\documentclass[a4paper, oneside, twocolumn, notitlepage, 10pt]{extarticle_ecoc}
\usepackage{ecoc}

\usepackage{amsmath,amssymb}
\usepackage{graphicx}
\usepackage{bm}
\usepackage[hidelinks]{hyperref}
\usepackage{siunitx}
\usepackage{svg}

\newcommand{\bA}{\boldsymbol{A}}
\newcommand{\bB}{\boldsymbol{B}}
\newcommand{\bM}{\boldsymbol{M}}
\newcommand{\bR}{\boldsymbol{R}}
\newcommand{\bP}{\boldsymbol{P}}
\newcommand{\bV}{\boldsymbol{V}}
\newcommand{\abs}[1]{\lvert #1 \rvert}
\newcommand{\norm}[1]{\lVert #1 \rVert}
\newcommand{\dd}{\mathrm{d}}
\newcommand{\jj}{\mathrm{j}}
\newcommand{\Leff}{L_\mathrm{eff}}
\DeclareMathOperator{\Real}{Re}

\begin{document}
\selectlanguage{english}    


\title{Regular Perturbation on the Group-Velocity Dispersion Parameter for Dual-Polarization Short-Reach Systems}%


\author{
    Dario Cellini\textsuperscript{(1)}, Vinícius Oliari\textsuperscript{(2)}, Erik Agrell\textsuperscript{(3)}, Marco Secondini\textsuperscript{(1)}, Gabriele Liga\textsuperscript{(4)}, Alex Alvarado\textsuperscript{(4)}
}

\maketitle                  


\begin{strip}
    \begin{author_descr}

        \textsuperscript{(1)} TeCIP Institute, Scuola Superiore Sant’Anna, Pisa, Italy,
        \textcolor{blue}{\uline{dario.cellini@santannapisa.it}}

        \textsuperscript{(2)} Nokia, San Jose, CA, United States

        \textsuperscript{(3)} Department of Electrical Engineering, Chalmers University of Technology, Gothenburg, Sweden

        \textsuperscript{(4)} Department of Electrical Engineering, Eindhoven University of Technology, Eindhoven, the Netherlands

    \end{author_descr}
\end{strip}

\renewcommand\footnotemark{}
\renewcommand\footnoterule{}


\begin{strip}
    \begin{ecoc_abstract}
        The Manakov equation governs the propagation of signals in dual-polarization systems. Its solution is usually approximated by regular perturbation on the nonlinear Kerr parameter. In this paper, we propose a novel regular perturbation on the group-velocity dispersion parameter for the Manakov equation. ©2026 The Author(s) 
    \end{ecoc_abstract}
\end{strip}


\section{Introduction}
\vspace{-.16cm}
Analytical models of fiber-optic propagation play an important role in the design and optimization of optical communication systems~\cite{Essiambre2010}.
They provide insights on the interplay between chromatic dispersion and nonlinearities, enable rapid performance evaluation avoiding computationally-expensive split-step Fourier method (SSFM) simulations, and serve as building blocks for new digital signal processing (DSP) algorithms such as digital nonlinearity compensation schemes. 
The most widely used analytical framework is the regular perturbation (RP) on the nonlinear coefficient~$\gamma$, which treats the fiber Kerr nonlinearity as a small perturbation of the linear (dispersive) propagation~\cite{Vannucci2002,Mecozzi2012}.
This approach accurately describes the physical signal propagation in the linear and pseudo-nonlinear transmission regimes, regardless of the accumulated chromatic dispersion in the link.

However, the accuracy of RP with respect to $\gamma$ degrades when nonlinear effects are strong, as is the case in systems operating at high launch powers, or with a small group-velocity dispersion (GVD) parameter $\beta_2$. To address this limitation, a complementary approach was proposed in~\cite{Oliari2019,Oliari2020}: the \emph{RP on the GVD parameter}~$\beta_2$, which treats GVD as a small perturbation and expresses the solution as a power series in $\beta_2$.
This approach works particularly well in regimes where nonlinear effects are beyond the reach of the RP on $\gamma$ and where GVD is limited. Examples include transmission in the O-band (near the zero-dispersion wavelength) or in non-zero dispersion-shifted fibers at moderately high powers.
In~\cite{Oliari2020}, the \emph{RP on} $\beta_2$ was developed for single-polarization systems and validated using the SSFM, demonstrating superior accuracy over RP on $\gamma$ in the weak-dispersion regime but showing its own limitations at high symbol rates.

A key application area for \emph{RP on} $\beta_2$ is the emerging O-band coherent data-center interconnect (DCI) segment, an operating scenario of growing relevance \cite{Berikaa2023}.
Standard single-mode fiber (SMF) has near-zero dispersion at \SI{1310}{\nm}. In this regime, the conventional RP on $\gamma$ rapidly loses accuracy with increasing power, while \emph{RP on} $\beta_2$ remains accurate as long as the accumulated dispersion is sufficiently small, as we will show.

In this paper, we extend the \emph{first-order RP on} $\beta_2$ of \cite{Oliari2020} to \emph{dual-polarization} systems by deriving the perturbative solution of the Manakov equation. We benchmark the proposed model against dual-polarization RP on $\gamma$, demonstrating the operational regions where each approach excels. 
To the best of our knowledge, this is the first analysis and validation of dual-polarization RP on $\beta_2$.

\section{RP on $\beta_2$ for Dual-Polarization Systems}

The propagation of dual-polarization signals through an optical fiber link is governed, in the absence of polarization mode dispersion, by the Manakov equation~\cite{Menyuk2006}
\begin{equation}
\frac{\partial \bA(t,z)}{\partial z} = -\jj\frac{\beta_2}{2}\frac{\partial^2 \bA(t,z)}{\partial t^2} + \jj\frac{8}{9}\gamma\, e^{-\alpha z}\norm{\bA}^2 \bA(t,z),
\label{eq:manakov}
\end{equation}
where the vector $\bA(t,z) = [{A}_\mathrm{x}(t,z), {A}_\mathrm{y}(t,z)]^\top$ is the Jones vector representing the propagating two-polarization power-normalized optical field, $\beta_2$ is the GVD parameter, $\gamma$ is the fiber Kerr coefficient, $\alpha$ is the attenuation coefficient, and $\norm{\bA}^2 = \abs{{A}_\mathrm{x}(t,z)}^2 + \abs{{A}_\mathrm{y}(t,z)}^2$ is the total instantaneous signal power.

The RP method~\cite{Vannucci2002} on the GVD parameter~\cite{Oliari2020} expands the solution of \eqref{eq:manakov} in powers of $\beta_2$. We apply it to the dual-polarization signal $\bA$ and truncate the power series to the first order, obtaining
\begin{equation}
\bA_\mathrm{RP}(t,z) = \bA_0(t,z) + \beta_2 \bA_1(t,z)
\label{eq:expansion}
\end{equation}
where $\bA_0$ and $\bA_1$ are determined in what follows.
Substituting ~\eqref{eq:expansion} in ~\eqref{eq:manakov} and setting $\beta_2 = 0$ yields the zeroth-order RP on $\beta_2$ solution
\begin{equation}
\bA_0(t,z) = \bA(t,0)\,e^{\jj\tfrac{8}{9}\gamma\norm{\bA(t,0)}^2 G(z)},
\label{eq:A0}
\end{equation}
where $G(z) = (1 - e^{-\alpha z})/\alpha$ is the fiber effective length.
This solution only includes a nonlinear phase rotation, depending on the overall input power $\norm{\bA(t,0)}^2$ of both polarizations, applied pointwise to the signal, it does not take into account the channel memory determined by chromatic dispersion.

Substituting~\eqref{eq:expansion} into~\eqref{eq:manakov} and grouping only the terms proportional to $\beta_2$ yields, after evaluating analytically the integrals that solve the resulting differential equation,
\begin{equation}
\bA_1(t,z) = \bB(t,z)\,e^{\jj\tfrac{8}{9}\gamma\norm{\bA(t,0)}^2 G(z)},
\label{eq:A1}
\end{equation}
where 
\begin{align}
\bB(t,&z) = -\bM(t) z + \bR(t)\, G_1(z) + \bP(t)\, G_2(z) \nonumber\\
& - \jj\frac{16}{9}\gamma\,\bA(t,0)\,\Real\!\big\{\langle\bA^*(t,0),\,\bV(t,z)\rangle\big\}
\label{eq:B}
\end{align}
carries the interaction between GVD and nonlinear effects. Here $\langle\mathbf{u},\mathbf{v}\rangle$ is the standard inner product for complex vector spaces, and
\begin{align}
\bM(t) &=\frac{\jj}{2}\,\frac{\dd^2 \bA(t,0)}{\dd t^2}, \nonumber\\
\bR(t) &= \frac{4}{9}\gamma\,\frac{\dd H(t)}{\dd t}\bA(t,0) + \frac{8}{9}\gamma\,H(t)\,\frac{\dd \bA(t,0)}{\dd t}, \nonumber\\
\bP(t) &=\jj\frac{32}{81}\,\gamma^2\,H^2(t)\,\bA(t,0)\,,
\label{eq:MRP}
\end{align}
with $H(t) \triangleq \dd \norm{\bA(t,0)}^2/\dd t$ being the time derivative of the instantaneous input power.
The function
\begin{align}
\bV(t,z) &= \bM(t)\big[G(z)z - G_1(z)\big] \nonumber\\
&\quad - \bR(t)\big[G(z)G_1(z) - G_2(z)\big] \nonumber\\
&\quad - \bP(t)\big[G(z)G_2(z) - G_3(z)\big]
\label{eq:V}
\end{align}
couples the two polarizations through the inner product in the r.h.s. of \eqref{eq:B}, and, finally,
\begin{equation}
G_n(z) = \int_0^z G^n(z')\,\dd z'
\label{eq:Gn}
\end{equation}
are the effective lengths to the power of $n$ integral functions, available in closed form~\cite[Eq. 27--29]{Oliari2020}.

\section{System Setup and Performance}

\begin{figure}[t]
\centering
\includegraphics[width=\columnwidth]{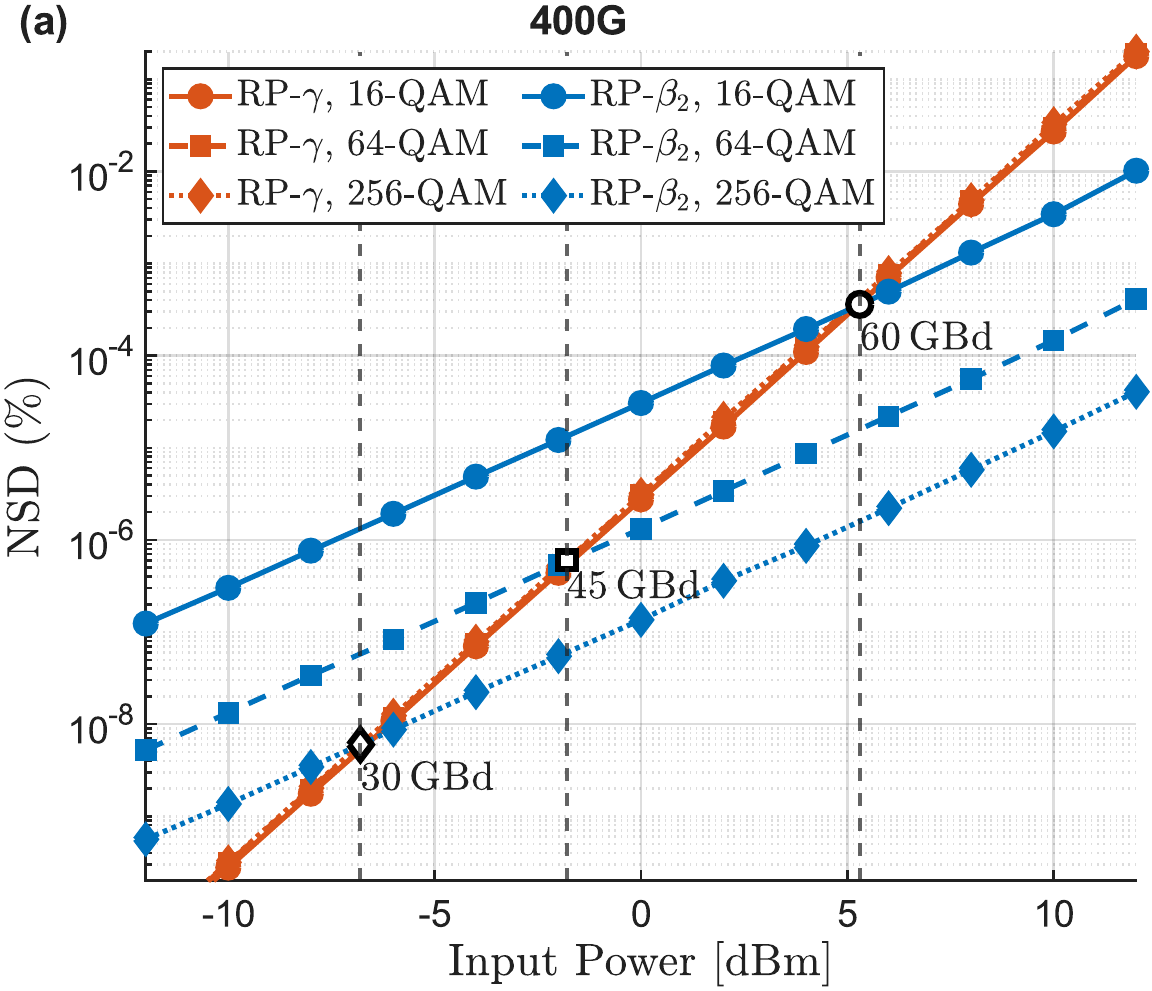}\\
\includegraphics[width=0.98\columnwidth]{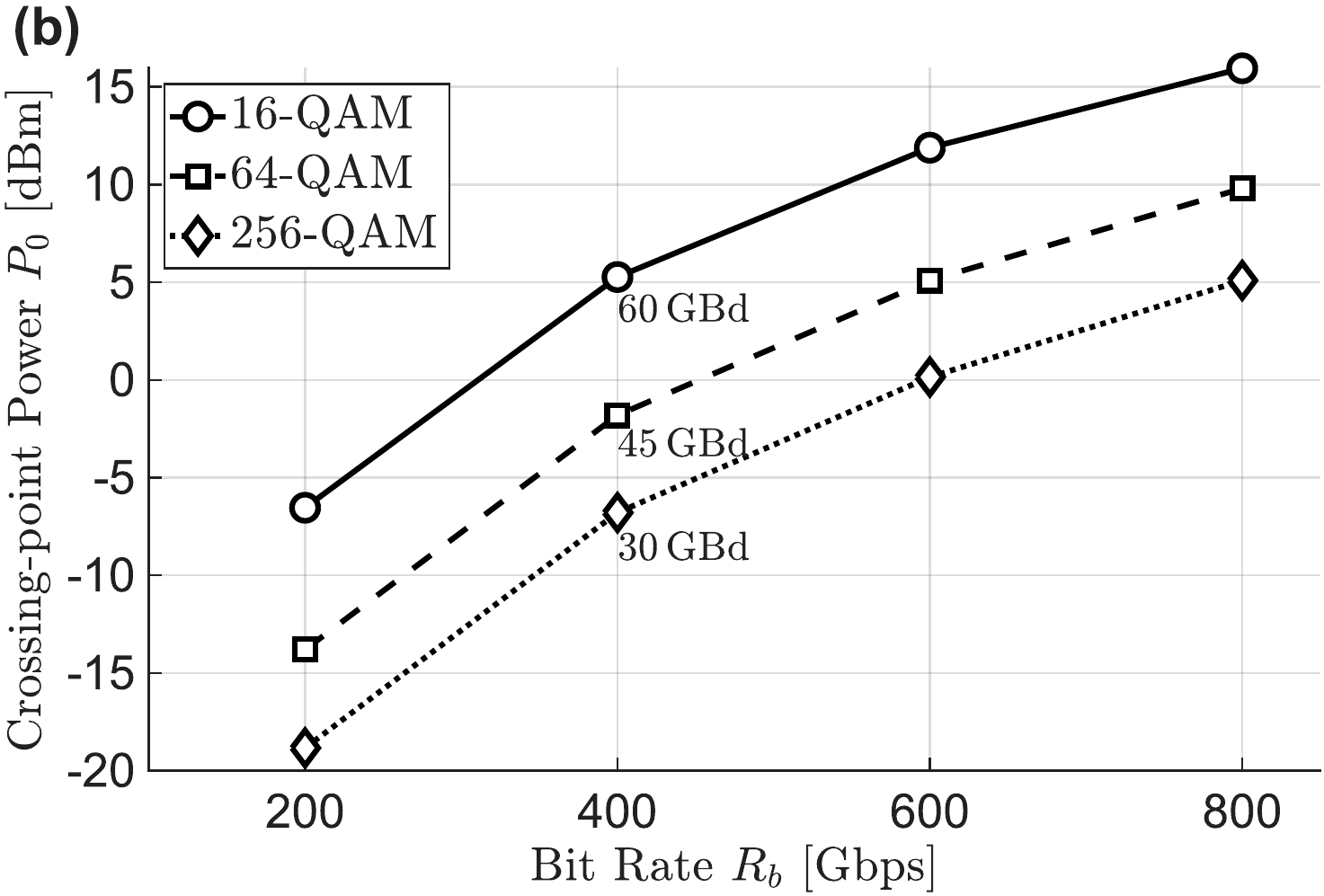}\\
\caption{(a)~NSD vs.\ total input power for RP-$\beta_2$ (blue) and RP-$\gamma$ (orange), at a fixed net bit rate of~$400~\mathrm{Gbps}$; (b)~Crossing-point power~$P_0$ vs.\ net bit rate.}
\label{fig:nsd_power}
\end{figure}

\begin{figure*}[!t]
\centering
\begin{minipage}[b]{0.3333\textwidth}
    \centering
    \includegraphics[width=\linewidth]{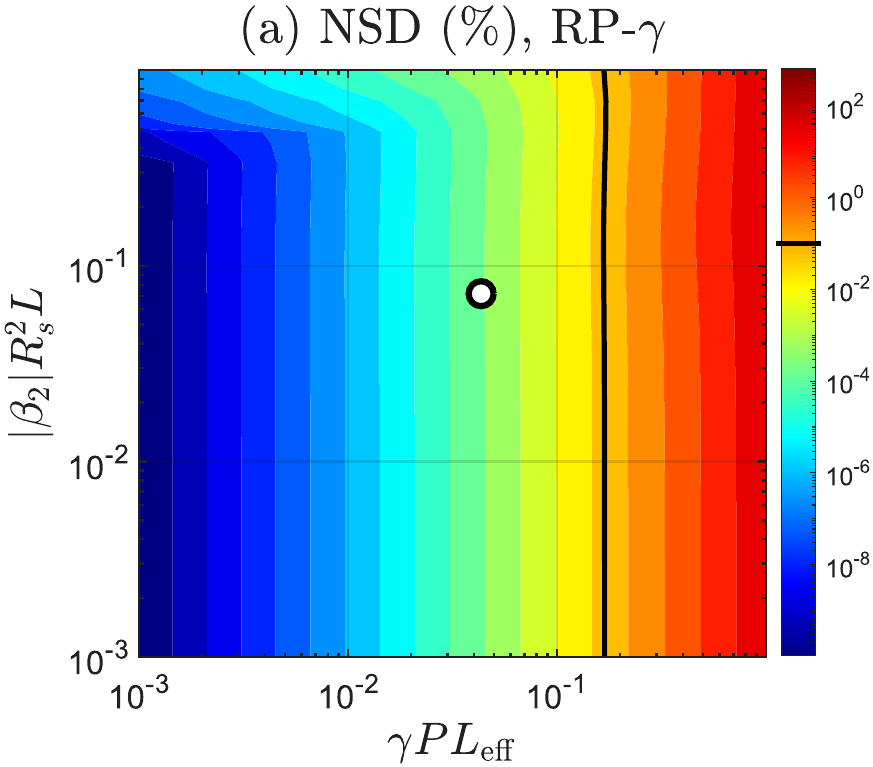}
\end{minipage}\hfill
\begin{minipage}[b]{0.3333\textwidth}
    \centering
    \includegraphics[width=\linewidth]{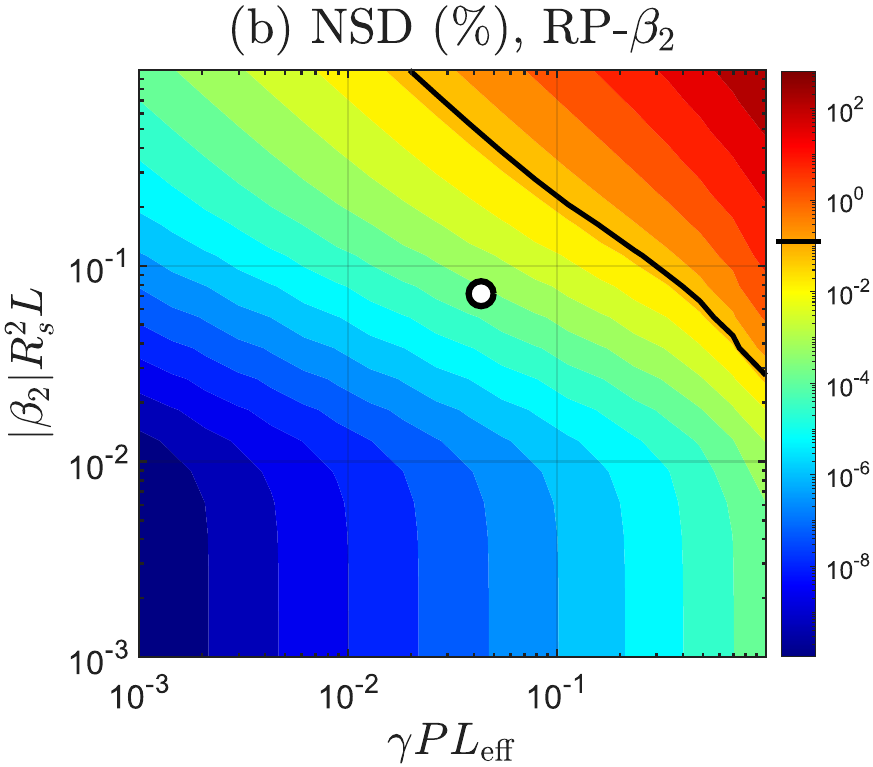}
\end{minipage}\hfill
\begin{minipage}[b]{0.3333\textwidth}
    \centering
    \includegraphics[width=\linewidth]{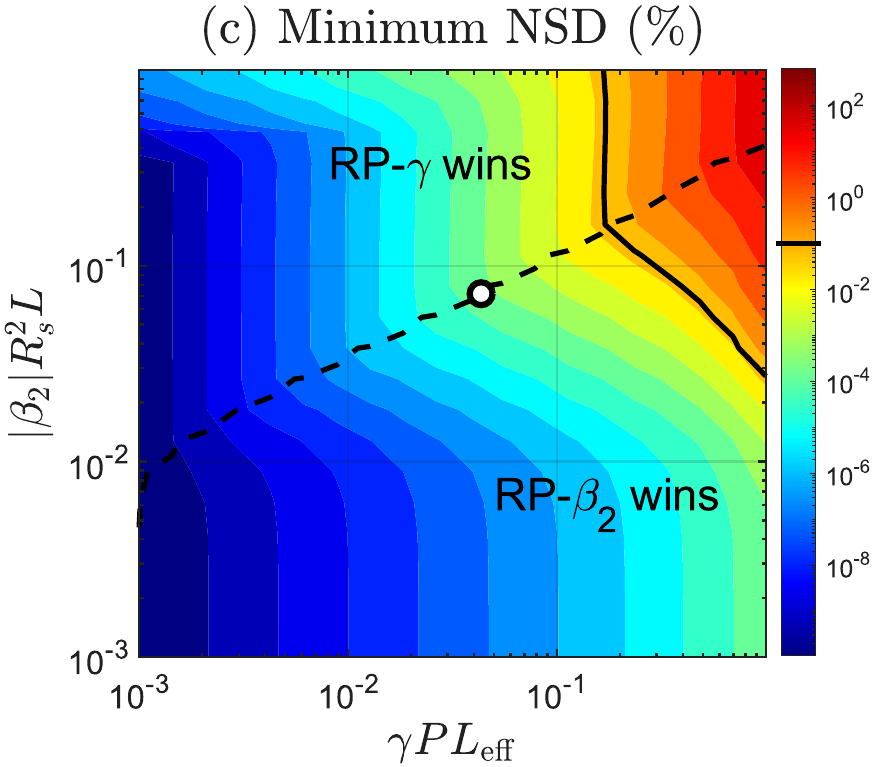}
\end{minipage}
\caption{NSD for (a)~RP-$\gamma$, (b)~RP-$\beta_2$, and (c)~the model-wise minimum error, obtained by always picking the more accurate of the two models. The solid black contour is the $\mathrm{NSD} = 0.1\,\%$ accuracy threshold, also marked on each colorbar. In panel~(c) the dashed line is made by the crossing-points of equal NSD, partitioning the plane into the RP-$\gamma$ and RP-$\beta_2$ favorable regions. The circle marker is the crossing-point of Fig.~\ref{fig:nsd_power}(a).}
\label{fig:contour}
\end{figure*}

We compare the first-order RP on ~$\beta_2$ in~\eqref{eq:expansion} (called RP-$\beta_2$) against the first-order RP on~$\gamma$ in their dual-polarization form~\cite{Tao2011} (called RP-$\gamma$). At the receiver, i.e., at $z=L$ (link length), the model error is measured, after applying chromatic dispersion compensation, by the normalized square deviation (NSD) defined as~\cite[Eq.~27]{Vannucci2002}
\begin{equation}
	\mathrm{NSD} = \frac{\int_{-\infty}^{+\infty}\norm{\bA_\mathrm{SSFM}(t,L) - \bA_\mathrm{RP}(t,L)}^2\,\dd t}{\int_{-\infty}^{+\infty}\norm{\bA_\mathrm{SSFM}(t,L)}^2\,\dd t},
	\label{eq:nsd}
\end{equation}
where $\bA_\mathrm{SSFM}$ represents the two continuous-time output waveforms of reference, computed using the symmetric SSFM with a sufficiently large number of steps. As a representative O-band DCI coherent link, we consider an SMF of $L = \SI{20}{\km}$, $\alpha = \SI{0.4}{\dB\per\km}$, $\gamma = \SI{1.4}{\per\W\per\km}$, $\beta_2 = \SI{-1}{\ps\squared\per\km}$, $\beta_3= \SI{0.0765}{\ps\cubed\per\km}$, dual-polarization transmission, and root-raised-cosine pulse shaping.

Fig.~\ref{fig:nsd_power}(a) depicts the NSD of the two RP models as a function of the total input power, at a fixed net bit rate of $400~\mathrm{Gbps}$ and for three modulation formats: 16-, 64- and 256-QAM.
Since net bit rate $R_b $ is fixed, the $M$-QAM symbol rate $R_s = R_b / (2 \log_2 M \times 0.83)$ is decreased from $60~\mathrm{GBd}$ for 16-QAM to $30~\mathrm{GBd}$ for 256-QAM, where the factor~$2$ accounts for the polarizations and $0.83$ accounts for a typical forward-error-correction overhead.
We observe that the NSD of RP-$\gamma$ grows steeply with power, and is virtually identical for the three modulation orders, showing that the model's accuracy is only sensitive to power but not to symbol rate. On the other hand, the NSD of RP-$\beta_2$ has a lower slope, and a different level for each modulation order, set by the dependence on symbol rate.
The crossing point of each pair of curves occurs at an input power $P_0$ that decreases as the modulation order increases. No additional dependence on modulation order has been observed in the error between the RP models and SSFM. 

Fig.~\ref{fig:nsd_power}(a) shows that above $P_0$, RP-$\beta_2$ is the more accurate model; below $P_0$, RP-$\gamma$ is preferred.
To translate this observation into practical design guidelines, Fig.~\ref{fig:nsd_power}(b) shows the crossing-point power $P_0$ as a function of net bit rate, for the same O-band SMF and link length, with 16-, 64-, and 256-QAM.
The value of $P_0$ increases with bit rate because a higher~$R_s$ enlarges the GVD-induced error of RP-$\beta_2$, shifting the balance toward RP-$\gamma$.
For a fixed bit rate, higher-order modulation formats require a lower $R_s$ and therefore yield a lower $P_0$, enlarging the region where RP-$\beta_2$ is advantageous.
For instance, at $400~\mathrm{Gbps}$ the crossing-point power moves from $5~\mathrm{dBm}$ for 16-QAM down to $-7~\mathrm{dBm}$ for 256-QAM, which is well within the typical operating range of O-band DCI systems.

The crossing-point power and the model accuracy depend on all system parameters.
To gain a unified view of the complementarity of the RP models, we introduce two dimensionless parameters:
\begin{equation}
\gamma P \Leff, \qquad \abs{\beta_2} R_s^2 L,
\label{eq:norm_params}
\end{equation}
where $P$ is the total input power and $\Leff = G(L) = (1-e^{-\alpha L})/\alpha$ is the effective length.
The quantity $\gamma P \Leff$ is the nonlinear phase rotation argument and represents the magnitude of nonlinearities: it controls the error of RP-$\gamma$, whose zeroth-order solution neglects nonlinearities. The quantity $\abs{\beta_2} R_s^2 L$ represents accumulated chromatic dispersion at the end of the link, and it controls the error of RP-$\beta_2$, whose zeroth-order solution neglects dispersion.

Fig.~\ref{fig:contour} reports the NSD for the first-order RP-$\gamma$, the first-order RP-$\beta_2$, and the model-wise minimum error, obtained by always picking the model with the lower NSD. The contour plots are obtained by sweeping $\gamma$ and $\abs{\beta_2}$ at fixed $P = 5~\mathrm{dBm}$, $R_s = 60~\mathrm{GBd}$, $L = \SI{20}{km}$, $\alpha = \SI{0.4}{\dB\per\km}$, for a dual-polarization 16-QAM format ($R_b=400~\mathrm{Gbps}$).
On every panel, the solid black contour marks the $\mathrm{NSD} = 0.1\,\%$ threshold under which a model is empirically defined accurate (also indicated by a horizontal black tick on each colorbar). 
The crossing-point power $P_0$ for the O-band 16-QAM scenario of Fig.~\ref{fig:nsd_power}(a) is marked on the plane with a circle marker.

While the NSD of RP-$\gamma$ (see Fig.~\ref{fig:contour}(a)) is nearly insensitive to $\abs{\beta_2} R_s^2 L$, the NSD of RP-$\beta_2$ (see Fig.~\ref{fig:contour}(b)) grows more rapidly along $\abs{\beta_2} R_s^2 L$ than along $\gamma P \Leff$ (the color stripes are denser along $\abs{\beta_2} R_s^2 L$). This confirms the complementarity of the two approaches.
Fig.~\ref{fig:contour}(c) shows in further detail the complementarity of the two models: the dashed line consists of the crossing points of equal accuracy of the models and partitions the plane into a region where RP-$\gamma$ is preferred (top-left) and a region where RP-$\beta_2$ is preferred (bottom-right).
The overall $0.1\,\%$ accuracy region is the union of the two individual accurate regions and covers a large portion of the plane.
In the top-right region, both first-order models are inaccurate, and thus, the choice of model does not matter.

\section{Conclusion}

We extended the first-order regular perturbation on~$\beta_2$ to 
dual-polarization systems and validated it against SSFM for a wide range 
of operating scenarios, including short-reach O-band links for next-generation 
DCIs. For a $\SI{20}{km}$-long link at $400~\mathrm{Gbps}$, 
RP-$\beta_2$ outperforms RP-$\gamma$ at launch powers above approximately 
$5~\mathrm{dBm}$ with 16-QAM, or lower when using higher-order modulation formats.
Through the analysis in terms of the dimensionless parameters $\gamma P \Leff$ and 
$\abs{\beta_2} R_s^2 L$, we confirmed that the accuracy of RP-$\beta_2$ 
depends mainly on $\abs{\beta_2} R_s^2 L$, while RP-$\gamma$ depends mostly on $\gamma P \Leff$. Furthermore, we 
studied their complementarity by identifying regions of applicability for 
each model.
By selecting the minimum NSD across the two models, $0.1\,\%$ NSD accuracy can be achieved over a vast region of the 
parameter space. This demonstrates that RP-$\beta_2$, used alongside RP-$\gamma$, 
provides a computationally efficient analytical tool for dual-polarization 
short-reach systems. The application of RP-$\beta_2$ to DSP algorithms 
such as digital backpropagation will be investigated in future work.

\clearpage

\section{Acknowledgements}
The work of D. Cellini and M. Secondini was partially supported by the European Union - Next GenerationEU under the Italian National Recovery and Resilience Plan (NRRP), Mission 4, Component 2, Investment 1.3, CUP J53C22003120001, partnership on “Telecommunications of the Future” (PE00000001 - program “RESTART”).


\printbibliography

\vspace{-4mm}

\end{document}